\documentclass[runningheads]{svmult}

\usepackage{makeidx}   % allows index generation
\usepackage{graphicx}  % standard LaTeX graphics tool
\usepackage{subeqnar}  % subnumbers individual equations
\usepackage{multicol}  % used for the two-column index
\usepackage{physprbb}  % modified textarea for proceedings,
\makeindex             % used for the subject index
\begin{document}
\title*{Globular Cluster Formation}
%
%
%\toctitle{Focusing of a Parallel Beam to Form a Point
%\protect\newline in the Particle Deflection Plane}
% allows explicit linebreak for the table of content
%
%
%\titlerunning{Focusing of a Parallel Beam}
% allows abbreviation of title, if the full title is too long
% to fit in the running head
%
\author{Keith M. Ashman}
%\and \inst{2}
%\and \inst{2}
%
%\authorrunning{Ivar Ekeland et al.}
% if there are more than two authors,
% please abbreviate author list for running head
%
%
\institute{University of Missouri--Kansas City, Department of Physics,
5110 Rockhill Road, Kansas City, MO 64110} 
%\and Universit\'{e} de Paris-Sud,
%     Laboratoire d'Analyse Num\'{e}rique,
%     B\^{a}timent 425,\\
%     F-91405 Orsay Cedex, France}

\maketitle              % typesets the title of the contribution

\begin{abstract}
The discovery of young globular clusters in merging galaxies and other
environments provides an opportunity to study directly the process of globular cluster
formation. Empirically it appears that globular cluster formation occurs
preferentially in regions in which star formation occurs at a high rate
and efficiency. Further, the interstellar medium in such regions is likely
to be at a higher pressure than less active star-forming environments.
An additional observational clue to the globular cluster formation process
is that young globular clusters have little or no mass-radius relationship.
In this paper I argue that high pressure and high star-formation
efficiency are responsible for current globular cluster formation. I suggest
that the precursors to globular clusters are molecular clouds and that 
the mass-radius relationship exhibited by such clouds is wiped out by a 
variable star formation efficiency.  
\end{abstract}

\section{Empirical Foundations}

Early models of globular cluster 
formation were largely motivated by two observational
results: Milky Way globular clusters are old and massive. Consequently, these
models tended to exploit physical conditions unique to the early universe
that might give rise to bound clusters of stars with masses 
around $10^5~M_{\odot}$. Over the last decade or so, observations of 
extragalactic globular cluster systems and the discovery of young globular
clusters have dramatically expanded the empirical basis of globular cluster
formation theories. Perhaps most importantly, young globular cluster systems
allow the formation process to be probed directly. 
In this section I describe the observations that are
useful in investigating and constraining the process of globular cluster formation.

\subsection{What do we know?}

The Milky Way globular cluster system is comprised of at least two
distinct populations (e.g.\ Armandroff and
Zinn 1988 and references therein). The more
numerous metal-poor clusters are distributed in a spherical halo, whereas
the metal-rich clusters have spatial and kinematic properties similar
to the bulge or thick disk. Despite the marked distinction in these properties
between the two populations, the mass distributions of the metal-poor and
metal-rich clusters are indistinguishable. Other spiral galaxies also
show evidence for
similar metal-rich and metal-poor populations of globular clusters, the
most compelling case being M31 (Ashman and Bird 1993;
Barmby et al 2001; Perrett et al 2002).

A similar metallicity dichotomy is now well-established in the globular cluster systems
of many elliptical galaxies (e.g., Kundu and Whitmore 2001; Larsen et al 2001). 
In the vast majority of cases, there
are also clear spatial distinctions between the populations with the
metal-rich clusters being more centrally concentrated than the metal-poor
ones. There are currently only a handful of detailed kinematic studies
of these systems (Zepf, these proceedings). In at least some of these studies
kinematic differences between the two globular cluster
populations have been demonstrated.
As in the case of spirals, the mass distributions of the two populations
of globular clusters within an elliptical
are indistinguishable. Further, the mass distributions of globular 
clusters in different galaxies are similar. 

Perhaps the most important development in understanding globular cluster 
formation was the discovery of young globular clusters in currently 
merging galaxies (e.g., the reviews of Schweizer 1998; Ashman and Zepf 1998). 
As discussed further below, this allows the
formation process to be studied directly rather than relying on extrapolations
based on observations of ancient globular clusters. More generally, this
discovery demonstrated that globular cluster formation is {\it not}
a process that is dependent on conditions unique to high redshifts.

The recent discovery of intermediate-aged globular clusters in a handful
of youngish ellipticals (see Goudfrooij in these proceedings and
references therein) provides a useful link between ancient globular
clusters and the very young objects in ongoing mergers. Of considerable
interest is the finding that the age of these intermediate systems are
consistent with the age of merger signatures in their host ellipticals.
This provides additional support to the idea that globular cluster formation
is not a uniquely cosmological phenomenon.

\subsection{What does it all mean?}

One of the traditional arguments against a pregalactic origin for 
globular clusters was the presence of color (interpreted as metallicity)
gradients in the globular cluster systems
of elliptical galaxies (Harris 1991). Clearly
a pregalactic origin is hard to reconcile with such an observation
since it requires higher metallicity clusters to preferentially adopt
smaller galactocentric distances.  However, the finding that these
color gradients are the result of two populations of globular
clusters with different spatial concentrations
complicates this conclusion. Since there is 
currently no definitive evidence for color gradients within the individual
populations, it is possible that one of the populations formed pregalactically,
later becoming associated with the parent galaxy through hierarchical
clustering. Equally, it is hard to avoid the conclusion that at least one
of the globular cluster populations of elliptical galaxies must have formed
within the galaxy itself. Again, if both populations were pregalactic there
is no obvious mechanism for generating the spatial (and kinematic)
differences between the two populations. Similar comments apply to
the metal-poor and metal-rich globular cluster systems of the Milky Way
and other spiral galaxies.

In order to explore this idea further, it is helpful to examine current
ideas on the formation of globular cluster systems. The presence of
metallicity bimodality in the globular cluster systems of ellipticals
was predicted in the context of the merger model (Ashman and Zepf 1992; see
also Zepf and Ashman 1993).
The metal-poor globular clusters are identified as those originally
in the halos of progenitor spirals, whereas the metal-rich ones are assumed
to form in the spiral-spiral merger that formed the elliptical. Thus
in this picture, the metal-rich clusters form with the elliptical, whereas
the metal-poor ones could have a pregalactic origin. In the dissipationless
hierarchical clustering scenario of C\^ot\'e et al (1998, 2002), metallicity
bimodality is attributed to the clustering of a large number of
galaxies and their associated globular cluster systems. The metal-poor
clusters are those associated with numerous dwarf galaxies, whereas
the metal-rich ones formed around the largest ``seed'' galaxy. Finally,
in the multiphase collapse model of Forbes et al (1997), both populations
of globular clusters form {\it in situ} within a collapsing elliptical,
with the metal-poor globular clusters forming first and the
metal-rich ones being produced in a secondary burst of star
formation. 

This discussion illustrates that in all extant models of globular cluster
{\it system} formation at least some globular clusters are formed within
galaxies. Indeed, in all cases, the metal-rich clusters are associated
with a significant star formation event in the parent galaxy.
Based on the observations outlined
above, this general result seems hard to dispute. Beasley et al (2002)
have recently studied this issue using a semi-analytic approach. They find
consistency with observation in schemes where metal-poor clusters form before
massive galaxies and metal-rich ones form in star-forming events associated
with massive galaxies such as mergers. The presence of
young and intermediate-aged globular clusters in mergers and merger 
remnants indicates that globular clusters {\it can} form in mergers,
but does not necessarily require that all globular clusters form in
such environments. Indeed, the globular cluster systems of dwarf
galaxies clearly did not form in major mergers. I will return to these
systems in Section~3.

\section{Globular cluster formation in mergers}

While not all globular clusters form in major mergers, the fact that some do
gives us an excellent starting point for investigating the globular
cluster formation process. This approach is made more attractive by
the evidence that globular cluster formation is rare in other star-forming
regions such as the disks of normal spirals. Larsen in these proceedings 
discusses ``young massive clusters'' in normal spiral disks. Whether these
objects are analogs of young globular clusters or whether they are more
diffuse objects is yet to be determined, but the critical issue is that
globular cluster formation is clearly more prevalent in regions where the
star formation rate is high such as merger-induced starbursts.

\subsection{The importance of pressure}

As noted by several authors (e.g., Elmegreen and Efremov 1997), the mass function
of Giant Molecular Clouds (GMCs) in the Milky Way and other nearby galaxies
and young globular clusters have similar slopes
when parameterized as power laws. Further, the slope of the mass function is
also consistent with that of old globular clusters at the high-mass end of
the distribution (e.g., Harris and Pudritz 1994). 

There are two
(possibly related) reasons why such GMCs do {\it not} produce a population of young
globular clusters in the Milky Way and similar environments. First, the
star formation efficiency in such GMCs is low. Consequently, the typical mass of
star clusters formed in these clouds is less than that of young globular
clusters. In fact, since the mass distribution is well-approximated by a power law,
high-mass clusters will still form from ordinary GMCs provided one has a sufficient
number of such clouds in a given galaxy. However, if
such massive clusters do form, they will not resemble globular clusters.
This is because the radii of GMCs in normal star-forming environments are much
greater than the characteristic radii of globular clusters.

One of the notable differences between the interstellar medium (ISM) in quiescent
disks and starbursts is that the pressure in the latter is inferred to be
much higher (e.g., Heckman et al 1993, 1990).
The relevance to the formation of dense star clusters is
immediately apparent. Clearly GMCs in a high-pressure environment will have 
higher densities and smaller radii than their counterparts in a galaxy
like the Milky Way. This is one reason why high pressure has been suggested
as a critical physical reason why globular clusters form in galaxy mergers
(Elmegreen and Efremov 1997; Ashman and Zepf 2001). 

To quantify this idea it is useful to employ the  Ebert-Bonner relations 
(Ebert 1955; Bonner
1956; see also Harris and Pudritz 1994; McLaughlin and Pudritz 1996)
for self-gravitating, pressure-bounded isothermal spheres: 
$$
M_c = \frac{ 3.45 }{ \gamma ^{3/2} } \frac{ \sigma ^4 }{ (G^3P_s)^{1/2} }
\eqno(1)
$$
$$
r_c = \frac{ 0.69 }{ \gamma ^{1/2} } \frac{ \sigma ^2 }{ (GP_s)^{1/2} }
\eqno(2)
$$
Here $M_c$ and $r_c$ refer to the mass and radius of the cloud,
$P_s$ is the cloud surface pressure,
$\gamma$ is a factor of order unity which is dependent on the
nature of the equilibrium, $\sigma$ is the one-dimensional
velocity dispersion within the cloud, and $G$ is the gravitational constant.

One can eliminate the velocity dipsersion from these expressions to obtain
a simple scaling relation:
$$
r_c \propto M_c^{1/2}P_s^{-1/4}
\eqno(3)
$$
If GMCs in the ISM of mergers are in equilibrium, the above relations and inferred
pressures in mergers imply that 
GMCs with masses of order $10^5~M_{\odot}$ have radii consistent with those of
young globular clusters. Along with the similarities in mass functions, this
result suggests that GMCs in high-pressure environments are at least plausible
progenitors to young globular clusters. 
At some level, this result is hardly surprising. Given that young globular 
clusters are found in merging galaxies, it is difficult to imagine any other
progenitor than dense molecular clouds. However, it is significant that
quantitatively the densities of such clouds in high-pressure environments
are consistent with the densities of young globular clusters. 

It is important to add that there is little direct information about the properties
of molecular clouds in these environments. It seems unlikely that globular cluster
progenitors are simply those GMCs originally in the disks of the merging spirals.
This is because the compression of such clouds when the surrounding warm ISM is
shock-heated is likely to cause cloud fragmentation before high densities are
reached (e.g., Jog and Solomon 1992; Jog and Das 1996). 
That is, the original GMCs of the spirals 
are unlikely to reach equilibrium with the high-pressure ISM before fragmentation.
It seems more probable that globular cluster progenitor clouds form within the
ISM of the merger once high pressures have been established.

\subsection{The strange case of the mass-radius relationship}

While the above considerations provide a simple framework for the
formation of globular clusters in mergers, there is one oddity that must be explained
if molecular clouds at high pressure are to be identified as globular cluster
progenitors. This is the observation that GMCs, at least in normal star-forming 
regions, have a mass-radius relation consistent with the Ebert-Bonner relations
given above [see equation (3)] 
whereas young globular clusters have a weak or non-existent relation
between mass and radius (Ashman and Zepf 2001). These
results have been established for GMCs in the Milky Way  (see the summary
of observations
given by Harris and Pudritz 1994), as well as M33, the LMC and
the SMC (Wilson and Scoville 1990; Johansson 1991; Rubio et al 1993). For young
globular clusters, observations of the galaxy merger
NGC~3256 indicate that there may be a weak correlation between
mass and radius (Zepf et al 1999), 
but one that is clearly much shallower
than the mass-radius relation of GMCs given in equation (3).
A similar weak or absent correlation between mass and radius is well-established
for the old globular clusters of the Milky Way  (van den Bergh
et al 1991; Djorgovski and Meylan 1994; Ashman and Zepf 1998) and also seems to hold
for the young star clusters in the LMC (van den Bergh 1991),
and for young star clusters in the Galaxy (e.g.\ Testi,
Palla and Natta 1999).

\subsection{A variable star formation efficiency}

Assuming that globular cluster progenitors are clouds in equilibrium, these
observations require that the original mass-radius relationship of such clouds
is wiped out during the globular cluster formation process. For this to occur
it is apparent that either the mass
or the radius (or both) of the final star clusters must differ from those
of the original clouds. One promising mechanism for producing such an effect
is that the star formation efficiency within clouds varies with mass and/or
radius. Much of the following discussion of this possibility follows the study
given in Ashman and Zepf (2001).

Let $\epsilon$ to be the star formation efficiency such that
$$
\epsilon = \frac{ M_* }{ M_c }
\eqno(4)
$$
where $M_*$ is the mass of the star cluster resulting from a cloud
mass $M_c$. To determine the mass-radius relation of
clusters, it is clear that we also need to be able to calculate
the final cluster radius, $r_*$, produced by a cloud with radius $r_c$.
In general, recently formed clusters are expected to undergo a phase
of expansion so that $r_* > r_c$.
We assume that a cloud fragments and produces an initial
cluster of radius $r_c$. If the gas loss is slow (i.e., it occurs
on timescales
longer than the cluster dynamical time) the product of mass and radius
is an adiabatic invariant. Under these conditions, the final cluster
radius is related to the star formation efficiency by:
$$
\frac{ r_* }{ r_c } \simeq \epsilon ^{-1}.
\eqno(5)
$$
(Hills 1980; Richstone and Potter 1982; Mathieu 1983). Thus lower star
formation efficiencies lead to greater expansion with sufficiently low
efficiencies producing unbound clusters. This expression
has recently been verified numerically by Geyer and Burkert (2001)
for the case of slow mass loss. For more rapid mass loss, these
authors find larger expansion rates at a given $\epsilon$, but for
$\epsilon < 0.4$ the clusters are unbound.

Ashman and Zepf (2001) investigated the consequences of a star formation
efficiency scaling with some power of cloud binding energy per unit mass, $M_c/r_c$:
$$
\epsilon \propto \left( \frac{ M_c }{ r_c } \right)^n
\eqno(6)
$$
Using equations (3) through (6) and some algebra one obtains a 
mass-radius relation for the resulting clusters:
$$
r_* \propto M_*^{(1-n)/(n+2)}
            P_s^{[1-2(n+1)]/[2(n+2)]}
\eqno(7)
$$
The weak or absent mass-radius correlation of globular clusters is
reproduced if the exponent on $M_*$ in equation (7) is close
to zero.  This occurs when $n \simeq 1$. For the specific case of
the young globular clusters in NGC~3256, Zepf et al (1999) found
$r_* \propto M_*^{0.1 \pm 0.1}$ (assuming a constant cluster mass-to-light ratio)
which is reproduced by $n \simeq 0.75 \pm 0.25$.
The value $n=1$ is interesting since it
corresponds to the case of a star formation efficiency which is directly
proportional to the binding energy per unit mass of the precursor
gas clouds. An increase in star formation
efficiency with velocity dispersion (which scales as the square root
of binding energy per unit mass, $n=0.5$) has been suggested by Elmegreen et al
(1993) and Elmegreen and Efremov (1997).

Throughout the above discussion it has been assumed that globular clusters form
from clouds in equilibrium, since it is this equilibrium that leads to the
cloud mass-radius relationship. It is important to note that there is currently
little information about the nature of molecular clouds in the ISM of merging
galaxies. There have been suggestions that the progenitor clouds to globular
clusters may not be in equilibrium prior to fragmentation 
(see McLaughlin in these proceedings and references therein). As far as I am
aware, the implications 
for the mass-radius relation of the resulting clusters have not been
investigated.

\subsection{Constraints on star formation efficiency variations}

One important aspect of this discussion is that independent considerations
place stringent constraints on variations in star formation efficiency. This
is because of the similarity of the mass function slopes of GMCs and young
globular clusters. Any star formation efficiency that includes a dependence on
mass will inevitably produce a cluster mass function with a different slope
to that of the progenitor clouds. To quantify this, consider the usual
parameterization of the mass spectrum of
clouds and clusters: 
$$
N(M_c)dM_c \propto M_c ^{-\beta} dM_c .
\eqno(8)
$$
$$
N(M_*)dM_* \propto M_* ^{-\alpha} dM_* ,
\eqno(9)
$$
These quantities can be related through the expression
$$
N(M_*)dM_* \propto N(M_c) \left( \frac{ dM_c }{ dM_* } \right) dM_*
\eqno(10)
$$

Ashman and Zepf (2001) showed that for a star formation efficiency given by
equation (6) this last expression implies a relation between the cloud and cluster
mass function slopes:
$$
\alpha = \frac{ 2 \beta + n }{ n + 2 } .
\eqno(11)
$$
Since $\alpha$ and $\beta$ are found observationally to be comparable, it is
apparent that the value of $n$ is constrained to be small. If, as argued by
Elemegreen and Falgarone (1996), $\beta=2$, then a typical observational value of
$\alpha$ of 1.8 implies $n=0.5$. Using these same values and the weak mass-radius
relation for young globular clusters in NGC~3256 leads to marginal consistency
with $n=1$. 

Note that in general this picture predicts that
$\alpha < \beta$.  Thus determinations of the mass distributions of clouds and
clusters have the potential to
refute or support this scenario. Unfortunately, the current uncertainties
in these quantities, as well as the fact that cluster and cloud mass
spectra are rarely derived for the same systems, do not allow a definitive
test of the scenario as yet. 
Future observations of the mass functions
of molecular clouds and globular clusters in the same system will address
this question. For example, ALMA will have the angular resolution and 
sensitivity to pin down the cloud mass spectrum in merging systems.
Interestingly, a dependence of star formation efficiency to
any positive power of cloud density, as is the case in Schmidt-type laws,
can already be ruled out (Ashman and Zepf 2001).

One further potential constraint on a variable star formation efficiency is that
for low enough values the resulting star clusters will be unbound. 
There are several complicating factors in
determining the mass-scale at which 
this occurs. For example, the center of a cloud might
produce stars with a sufficiently high efficiency to form a bound cluster, but
the global star formation efficiency (relevant to the arguments above) might
be low. More generally, the lowest star formation efficiency that can produce a bound 
cluster is still a debated question. 
It is also worth noting that in the current picture it is the low-mass
clusters that have the lowest star formation efficiencies and thus are most
likely to be unbound. The dissolution of these clusters in this manner may be
relevant to low-mass cluster destruction in general, which is required if
the young globular cluster mass function is to evolve into that of old
globular clusters (e.g., Fall and Rees 1977; Murali and Weinberg 1996; Gnedin and
Ostriker 1997; Vesperini 1997 and in these proceedings).

\subsection{The connection between pressure and star formation efficiency}

A critical question in this discussion is {\it why} there might be a relationship
between star formation efficiency and the surface pressure of molecular clouds.
First, it is important to note that observationally the star formation efficiency
in merger-induced starbursts is high. Several authors have attributed this 
high star formation efficiency to the high pressure in such environments 
(e.g., Jog and Solomon 1992; Jog and Das 1996; Elmegreen and Efremov 1997).
It is exactly the high ambient pressure, of course, that produces clouds
with a high binding energy. 

More generally, there are plausible reasons why star formation efficiency might
depend on the binding energy per unit mass of clouds
(see also Elemegreen et al 1993; Elmegreen and Efremov 1997).
To a first approximation, the disruptive energy input from massive
stars will be proportional to the number of such stars and thus the
mass of the cloud, hence the normalization of binding energy to unit mass.
It seems likely that clouds with a higher
binding energy will be less affected by such disruption and therefore
convert a higher fraction of their gas mass into stars. 
It is important to distinguish in this context between global and local effects.
Clearly binding energy considerations are central to establishing whether
a young cluster will remain bound at all. In terms of a rationale for a star
formation efficiency dependent on binding energy, the issue is that 
{\it local} feedback
effects are likely to be more important (and therefore more likely to suppress further
star formation) in clouds of higher binding energy.

\subsection{More on cloud fragmentation}

In the above discussion no attempt has been made to address the details of the
cloud fragmentation process. Further, the Ebert-Bonner relations that underpin
the scaling arguments refer to the mean properties of clouds. Consequently, the
potentially important issue of the density profiles of clouds
 is not addressed in the above
approach. It seems inevitable that any understanding of fragmentation must include
a study of how fragmentation and subsequent feedback processes occur locally
within clouds, and thus on the density profile of clouds. Many of these issues
are discussed in the comprehensive review by Elmegreen (2002; see also
McLaughlin in these proceedings).

Of possible relevance to the elimination of the mass-radius relationship is the fact
that fragmentation is dependent on the equation of state of the cloud
(see, for example, Li in these proceedings). This can be understood using standard
Jeans mass arguments. If the equation of state is expressed using the usual
polytropic form:
$$
P \propto \rho ^{\gamma}
\eqno(12)
$$
it follows that the Jeans mass can be written
$$
M_J \propto \rho ^{3/2(\gamma - 4/3)}
\eqno(13)
$$
One expects fragmentation to proceed if cloud contraction, and thus an increase in
cloud density, leads to a decrease in the Jeans mass. Clearly this occurs when 
$\gamma < 4/3$. Simulations by Li (these proceedings; see also Spaans and Silk 2000
and references therein) support this expectation.

Of interest in this context is the fact that the critical value of
$\gamma = 4/3$ also corresponds to clouds with no mass-radius relationship. 
This is because such a value implies that cloud
mass is independent of density, as is apparent from
equation (13). If the progenitor clouds of globular clusters initially
have $\gamma > 4/3$ with this value subsequently decreasing, it is
possible that fragmentation begins once the value of 4/3 is reached. 
Consequentky, the
resulting clusters would likely have no mass-radius relationship. 
The critical issue is therefore 
whether molecular clouds in starburst enviroments
are likely to experience this kind of evolution. Unfortunately,
there is currently no clear consensus on the equation of state of 
molecular clouds and
related stability issues even in systems where these objects are well 
studied (e.g., McLaughlin and Pudritz 1996; McKee and Holliman 1999;
Curry and McKee 2000 and references therein).

Despite this uncertainty, there are general
considerations that suggest molecular clouds in starbursts may be initially
characterized by large values of $\gamma$ (assuming a single polytropic
index is adequate at all; see Curry and McKee 2000). As argued earlier,
clouds that produce globular clusters in such environments probably formed
{\it after} the initial shock-heating of the ISM, since pre-exisiting
clouds would fragment before reaching the densities typical of globular
clusters. Consequently, the progenitor clouds will be in an environment
with a significant radiation field. The resulting heating of clouds will
tend to push $\gamma$ to large values. Spaans and Silk (2000) have
investigated this issue and note that opaque dust in starbursts 
is an additional factor that
tends to lead to large values of $\gamma$. These authors also find that
$\gamma$ subsequently decreases towards unity in such enviroments. It therefore seems 
at least plausible that there is a physical connection between the
critical value of $\gamma$ required for cloud fragmentation and the
absence of a mass-radius relation in globular clusters.

\section{The origin of metal-poor clusters}

While there is compelling
 evidence that metal-rich globular clusters formed during major
star-forming events within their parent galaxies, the origin of metal-poor
clusters is less clear. Empirically, metal-poor clusters are found in a 
wide range of environments from the faintest dwarfs to the halos of massive
spirals and ellipticals. This ubiquity suggests that the metal-poor
globular clusters may have a pregalactic origin.
However, it seems unlikely that such globular
clusters represent the first bound structures to form in the universe,
as originally envisaged by Peebles and Dicke (1968). The well-known problem
is that such cosmological structures are expected to be surrounded
by dark matter halos, whereas observations of metal-poor globular
clusters in the Milky Way rule out the presence of such halos (e.g., Moore 1996).

One interesting possibility for resurrecting a pregalactic origin has
recently been proposed by Bromm and Clarke (2002). These authors
have carried out numerical simulations of structure formation within 
dwarf galaxies at early epochs. They suggest that gas in
dark matter ``subhalos'' within such dwarfs fragments to form the stellar
component of globular clusters and that subsequent violent relaxation of
the dwarf galaxy itself wipes
out the individual subhalos around the globular clusters. The mass of
globular clusters in this picture is thus effectively set by the 
mass of the subhalos and the ratio of gas to dark matter. This differs
from the original Peebles and Dicke (1968) scenario in the sense that
globular cluster formation occurs within a larger bound system, but it
does connect at least some globular cluster formation with cosmological
conditions through the dark matter mass spectrum.

To some extent, the picture of Bromm and Clarke (2002) is similar to
other ideas in which the sites of metal-poor globular clusters are
larger objects with masses around $10^8 M_{\odot}$. There are several
motivations for such a view. For instance, there is evidence that the halo
of the Milky Way was assembled from ``Searle-Zinn'' 
sub-galactic fragments (e.g., Searle and Zinn 1978). 
Further, such a picture is consistent with
succesful models of the formation of cosmological structure.
Current dwarf galaxies may be the surviving remnants of such objects
(see, however, the caveats presented by Santos in these proceedings).
Hierarchical clustering of some of these fragments into larger galaxies
leads to metal-poor globular clusters in the halos of spirals and
ellipticals. Along similar lines, 
Harris and Pudritz (1994; also McLaughlin and Pudritz 1996) 
presented a globular cluster formation model
in which the sites of formation are ``Super Giant Molecular Clouds''
(SGMCs). While it may be difficult to produce such massive clouds
in current galaxy mergers (e.g., Ashman and Zepf 1998), such SGMCs are similar to the
Searle-Zinn fragments discussed above.

In Section~2 I argued that high pressure is a necessary condition for
globular cluster formation. It is therefore of some interest to
establish whether high-pressure conditions might have existed in
sub-galactic fragments at earlier epochs. Steve Zepf and I are currently
investigating this question. Our preliminary results suggest that
high pressure conditions can be achieved in such systems through feedback
processes associated with massive stars. However, the shallow potential
wells of these systems mean that the gas is unlikely to remain bound to
the fragments for long (see also Dekel and Silk 1986). The implications
for globular cluster formation are currently being investigated.

A critical observational issue at the center of understanding the
formation of metal-poor clusters is the uniformity of the metallicity
of these objects. If all metal-poor globular cluster systems have
similar mean metalicities, it suggests that all metal-poor clusters
formed in similar environments. In this case, pregalactic formation in sub-galactic
fragments is an attractive possibility (e.g., Ashman and Zepf 1992;
Ashman and Bird 1993). Some variation
in mean metallicity would not rule out this option provided it did not
correlate with properties of the current parent galaxy. This is because
there is evidence that the mean metallicity of the globular cluster systems
of dwarf galaxies increases with galaxy luminosity (e.g., Lotz, these
proceedings and references therein). 

\section{Conclusions}

While there are still many aspects of globular cluster formation that
are poorly understood, there does appear to have been significant
progress over the last few years. Specifically, the discovery of
young globular clusters in ongoing mergers has provided an empirical
basis to the study of globular cluster formation. One notable result
is that there is no longer any need to invoke conditions unique to
the early universe in order to explain the origin of these objects.
Indeed, the approach to understanding globular cluster formation 
advocated in this paper is to first understand why globular clusters
form in such abundance in regions where the star-formation rate is high.

The realization that globular cluster formation is not a process unique
to the early universe has also made globular clusters themselves less
unique. Current evidence is consistent with the view that 
all star clusters are fundamentally similar and that globular clusters
represent one end of the star cluster spectrum. This view is reinforced
by the finding that both open and globular clusters share the curious weak
or absent correlation between mass and radius. While this state of affairs
may produce some semantic issues, such as how to define a globular cluster,
it also offers the exciting possibility of a unified approach to the
formation of all star clusters.

\bigskip

\bigskip

The work described in this paper would not have been possible without the
many collaborators with whom I have worked on these topics. My 
collaboration with Steve Zepf is particularly notable, both for its
productivity and longevity. 
 Some of the ideas in this paper were improved  
by many stimulating conversations with other workshop participants including
Henny Lamers, Yuexing Li, Arunav Kundu, Dean McLaughlin and Enrico Vesperini.
This work was supported in part by NASA Astrophysics Theory grant
NAG5-11320.

\def\ref{\par\noindent\hangindent 20pt}

%\begin{thebibliography}{8.}
%\addcontentsline{toc}{section}{References}

\section*{References}

\ref
Armandroff, T.E., \& Zinn, R. 1988, AJ, 96, 92

\ref
Ashman, K.M., \& Bird, C.M. 1993, AJ, 106, 2281

\ref
Ashman, K.M., \& Zepf, S.E. 1992, ApJ, 384, 50

\ref
Ashman, K.M., \& Zepf, S.E. 1998, Globular Cluster Systems,
Cambridge University Press.

\ref
Ashman, K.M., \& Zepf, S.E. 2001, AJ, 122, 1888

\ref
Barmby, P., Huchra, J.P., \& Brodie, J.P. 2001, AJ, 121, 1482

\ref
Beasley, M.A., Baugh, C.M., Forbes, D.A., Sharples, R.M., \& Frenk, C.S.
2002, MNRAS, 333, 383

\ref
Bonner, W.B. 1956, MNRAS, 116, 351

\ref
Bromm, V., \& Clarke, C.J. 2002, ApJLett, 566, L1

\ref
C\^ot\'e, P., Marzke, R.O., \& West, M.J. 1998, ApJ 501, 554

\ref
C\^ot\'e, P., West, M.J., \& Marzke, R.O. 2002, ApJ, 567, 853

\ref
Curry, C.L., \& McKee, C.F. 2000, ApJ, 528, 734

\ref
Dekel, A., \& Silk, J. 1986, ApJ, 303, 39

\ref
Djorgovki, S.G. \& Meylan, G. 1994, AJ, 108, 1292

\ref
Ebert, R. 1955, Z. Astrophys, 37, 222

\ref
Elmegreen, B.G. 2002, ApJ, 577, 206

\ref
Elmegreen, B.G., \& Efremov, Y.N. 1997, ApJ, 480, 235

\ref
Elmegreen, B.G., \& Falgarone, E. 1996, ApJ, 471, 816

\ref
Elmegreen, B.G., Kaufman, M., \& Thomasson, M. 1993, ApJ, 412, 90

\ref
Fall, S.M., \& Rees, M.J. 1977, MNRAS, 181, 37

\ref
Forbes, D.A., Brodie, J.P., \& Grillmair, C.J. 1997, AJ, 113, 1652

\ref
Geyer, M.P., \& Burkert, A. 2001, MNRAS, 323, 988

\ref
Gnedin, O.Y., \& Ostriker, J.P. 1997, ApJ, 474, 223

\ref
Harris, W.E. 1991, ARAA, 29, 543

\ref
Harris, W.E., \& Pudritz, R.E. 1994, ApJ, 429, 177

\ref
Heckman, T.M., Armus, L., \& Miley, G.K. 1990, ApJS, 74, 833

\ref
Heckman, T.M., Lehnert, M.D., Armus, L. 1993, in The Environment
and Evolution of Galaxies, ed. J.M. Shull \& Thronson, H.A., Jr.\
(Dordrecht: Kluwer), 455

\ref
Hills, J.G. 1980, ApJ, 225, 986

\ref
Jog, C., \& Das, M. 1996, ApJ, 473, 797

\ref
Jog, C., \& Solomon, P. 1992, ApJ, 387, 152

\ref
Johannson, L.E.B. 1991, in IAU Symp.\ 146, Dynamics of
        Galaxies and Their Molecular Cloud Distributions, ed. F. Combes \&
        F. Casoli, (Dordrecht: Kluwer), 1

\ref
Kundu, A., \& Whitmore, B.C. 2001, AJ , 121, 2950

\ref
Larsen, S.S., Brodie, J.P., Huchra, J.P., Forbes, D A., \&
Grillmair, C. J., 2001, AJ, 121, 2974

\ref
Mathieu, R.D. 1983, ApJ, 267, L97

\ref
McKee, C.F., \& Holliman, J.H. 1999, ApJ, 522, 313

\ref
McLaughlin, D.E., \& Pudritz, R.E. 1996, ApJ, 457, 578

\ref
Moore, B. 1996, ApJLett, 416, L13

\ref
Murali, C., \& Weinberg, D. 1996, MNRAS, 288, 767

\ref
Peebles, P.J.E., \& Dicke, R.H. 1968, ApJ, 154, 891

\ref
Perret, K.M., Bridges, T.J., Hanes, D.A., Irwin, M.J., Brodie, J.P.,
Carter, D., Huchra, J.P., \& Watson, F.G. 2002, AJ, 123, 2490

\ref
Richstone, D.O., \& Potter, M.D. 1982, ApJ, 254, 451

\ref
Rubio, M., Lequeux, J., \& Boulanger, F. 1993, A\&A, 271, 9

\ref
Schweizer, F. 1998, in Galaxies: Interactions and Induced
Star Formation, Saas-Fee Course 26, eds.\ R.C. Kennicutt et al
(Springer: Berlin), 105

\ref
Spaans, M., \& Silk, J. 2000, ApJ, 538, 115

\ref
Testi, L., Palla, F., \& Natta, A. 1999, A\&A, 342, 515

\ref
van den Bergh, S. 1991, ApJ, 369, 1

\ref
van den Bergh, S., Morbey, C.L., \& Padzer, J. 1991, ApJ, 375, 594

\ref
Vesperini, E. 1997, MNRAS, 287, 915

\ref
Wilson. C.D., \& Scoville, N. 1990, ApJ, 363, 435

\ref
Zepf, S.E., \& Ashman, K.M. 1993, MNRAS, 264, 611

\ref
Zepf, S.E., Ashman, K.M., English, J., Freeman, K.C.,
\& Sharples, R.M. 1999, AJ, 118, 752

%\end{thebibliography}

%INDEX%%%%%%%%%%%%%%%%%%%%%%%%%%%%%%%%%%%%%%%%%%%%%%%%%%%%%%%%%%%%%%%
% Please check with the editor of your book whether he plans to
% include a "mutual" subject index - if so, please code your entries
% in the standard syntax. For your own purposes you may print your
% "personal" index by using the following commands:
%
%\clearpage
%\addcontentsline{toc}{section}{Index}
%\flushbottom
%\printindex
%%%%%%%%%%%%%%%%%%%%%%%%%%%%%%%%%%%%%%%%%%%%%%%%%%%%%%%%%%%%%%%%%%%%%

\end{document}